\documentclass[prb,preprint,showpacs,showkeys,amsmath,amssymb,superscriptaddress]{revtex4}
\usepackage{amsmath,graphicx,latexsym,times,color}
\usepackage{setspace}
\usepackage{hyperref}
\usepackage{array}
\usepackage{slashbox}
\usepackage{titlesec}

\begin{document}

\title{Magnetocrystalline anisotropy energy of Fe$(001)$, Fe$(110)$  slabs and nanoclusters: a detailed local analysis within a tight-binding model}

\author{Dongzhe Li}
\affiliation{IRAMIS, SPCSI, CEA Saclay, 91191 Gif-sur-Yvette Cedex, France}

\author{Alexander Smogunov}
\affiliation{IRAMIS, SPCSI, CEA Saclay, 91191 Gif-sur-Yvette Cedex, France}

\author{Cyrille Barreteau}
\email{cyrille.barreteau@cea.fr}
\affiliation{IRAMIS, SPCSI, CEA Saclay, 91191 Gif-sur-Yvette Cedex, France}

\author{Fran\c cois Ducastelle}
\affiliation{Laboratoire d’Etude des Microstructures, ONERA-CNRS, BP 72, 92322 Châtillon Cedex, France}

\author{Daniel Spanjaard}
\affiliation{Laboratoire de Physique des Solides, Université Paris Sud, Batiment 510, F-91405 Orsay, France}

\date{\today}

\begin{abstract}
We report tight-binding (TB) calculations of magnetocrystalline anisotropy energy (MAE)  of Iron slabs and nanoclusters with a particuler focus on local analysis. After clarifying various concepts and formulations for the determination of MAE, we apply our realistic  TB model to the analysis of the magnetic anisotropy of Fe$(001)$, Fe$(110)$ slabs and of two large Fe clusters with $(001)$ and $(110)$ facets only: a truncated pyramid and a truncated bipyramid containg 620 and 1096 atoms, respectively. It is shown that the MAE of slabs originates mainly from outer layers, a small contribution from the bulk gives rise, however,  to an oscillatory behavior for large thicknesses. Interestingly, the MAE of the nanoclusters considered is almost solely due to $(001)$ facets and the base perimeter of the pyramid. We believe that this fact could be used to efficiently control the anisotropy of Iron nanoparticles and could also have consequences on their spin dynamics.
\end{abstract}



\newcommand{\Ea}{\ensuremath{E_a}}
\newcommand{\Eb}{\ensuremath{E_b}}
\newcommand{\Ec}{\ensuremath{E_c}}
\newcommand{\Ed}{\ensuremath{E_d}}
\newcommand{\Ee}{\ensuremath{E_e}}

\newcommand{\eV}{\ensuremath{\,eV}}

\maketitle

\section{INTRODUCTION}

The magnetic anisotropy which is characterized by the dependence of the energy of a magnetic system on the orientation of its magnetization is a quantity of central importance. The orientation corresponding to the minimum of energy (so called easy axis)  determines the magnetization direction at low temperature. The width of a domain wall  directely related to the strength of the anisotropy. The spin dynamics is also very much influenced by the shape of the magnetic energy landscape. For example the thermal stability of small nanoparticles with respect to magnetization reversal is controled by the height of the energy barrier to overcome during the switching process of the magnetization. The development of materials with large uniaxial anisotropy is very useful for technological applications such as high density magnetic recording or memory devices.  For example the storage unit can be made up by metallic grains and higher storage densities is achieved by reducing the magnetic grains down to nanoscale.

The origin of magnetic anisotropy is twofold: magnetostatic interaction and spin-orbit coupling\cite{Bruno1989}. The first one gives rise of the so-called “shape” anisotropy since it depends on the shape of the sample while the second is responsible for the overall magnetocrystalline anisotropy  energy (MAE). The shape anisotropy of “classical” origin needs not to be included in electronic structure calculation and can be added a posteriori by summing all pairs of magnetic dipole-dipole interaction energies. In thin films it favors in plane magnetization and is proportional to the thickness of the film and generally dominates for thick enough films. It will not be considered hereafter since it behaves almost linearly with the film thickness and cannot be at the origin of any MAE oscillations. The MAE on the contrary is a purely quantum effect and has a more complex behaviour. Its value per atom  is usually extremely small in bulk ($\mu$ eV) but can get larger in ultrathin films, multilayers or nanostructures.

Due to the smallness of the energy differences in play,  the determination of MAE still remains numerically delicate. However, there now exists a vast body of  reasearch devoted to the calculations of MAE in monolayers\cite{Gay1986,Bruno1989}, multilayers\cite{Szunyogh1995,Szunyogh1997,Ujfalussy1996,Ujfalussy1996a}, thin films\cite{Cinal1994}, clusters\cite{Pastor1995, Nicolas2006} or nanowires\cite{Autes2006} systems with ab-initio as well as tight-binding electronic structure methods.  
Technically several approaches have been developed for the determination of the MAE. The brute force method  consists in performing self consistent calculations for various orientations of the magnetization. This approach although straightforward is the most computationnally demanding since it usually necessitates a long self-consistent loop that implies the diagonalization of large matrices. Rather early it was recognized that small changes of the total energy could be related to the changes of the eigenvalues of the Hamiltonian. This is the so-called Force Theorem\cite{Weinert1985,Daalderop1990,Wang1996} that is very well suited to the calculation of MAE and has been used extensively in the past. Besides its computational efficiency it is also very stable numerically. Several works are also based on a perturbative treatment that consists in writing to second order the energy correction due to the spin-orbit Hamiltonian treated as a perturbation\cite{Bruno1989,Cinal1994}. Finally to get a basic understanding of the underlying physical phenomena it is very convenient to write the total MAE as a sum of contributions arising from atoms with bulk-like environment and from atoms with lower local symmetry such as surface/interfaces atoms.  In line with the various methods exposed above there are also many different ways to decompose the total MAE into atomic contributions and it is not always clear whether or not they are all  valid.

In this paper we wish to propose a comprehensive overview that will clarify the different aproaches to calculate the magnetocrystalline anisotropy with a particular emphasis on the atomic site decomposition of MAE and application to Iron slabs and clusters.  We will first describe in Sec.\ref{Sec:Method} our tight-binding method used thoughout the paper and then present in detail two alternative versions of the Force Theorem (FT)  widely used in the litterature.  We will basically show that even though these two versions of the FT are equivalent in terms of total energy variatons, the so-called grand-canonical FT is the most suited to define local quantities. In Sec. \ref {Sec:MAEslabs} we illustrate theses concepts on the case of Fe$(001)$ and Fe$(110)$ slabs that behave very differently. In particular $(001)$ surface favors out of plane magnetization while the easy axis of $(110)$ slabs is in-plane and a  with a smaller anisotropy. This has important consequences on the MAE of nanoclusters that are investigated in the Sec. \ref{Sec:MAEnanoclusters}. Finally in Sec.\ref{Sec:conclusion} we will draw the main conclusions of this work.

\section{Method and computational details}

 \label{Sec:Method}
\subsection{Magnetic Tight-Binding model}
In the following most of our calculations are based on an efficient tight-binding (TB) model including magnetism and spin-orbit interaction that has been described in several publications \cite{Autes2006, Barreteau2012}. We will just briefly recall its main ingredients. The Hamiltonian in a non-orthogonal $s,p,d$ pseudo atomic basis is written as a sum of 4 terms $H=H_{\text{TB}}+H_{\text{LCN}}+H_{\text{Stoner}}+H_{\text{SO}}$. Where $H_{\text{TB}}$ is a standard "non-magnetic" TB hamiltonian which form is very similar to the one introduced by Mehl and Papaconstantopoulos\cite{Mehl1996}, $H_{\text{LCN}}$ is a term ensuring local charge neutrality, $H_{\text{Stoner}}$  a Stoner-like contribution that controls the spin magnetization and $H_{\text{SO}}$ coresponds to spin-orbit interaction.   

Within this model the total energy of the system should be corrected by the so-called double counting terms arising from electron-electron interaction introduced by local charge neutrality and Stoner terms. The total energy then takes the form as follows:

\begin{equation}
E_{\text{tot}}=E_{\text{b}}-E_{\text{dc}}
\end{equation}

where $E_{\text{b}}=\sum_{\alpha} f_{\alpha} \epsilon_{\alpha}$ is the band energy ($ f_{\alpha}=f(\epsilon_{\alpha})$ being the Fermi-Dirac occupation of state $\alpha$ and corresponding eigenvalue $\epsilon_{\alpha}$). The expression of the double counting term is given by:

\begin{equation}
E_{\rm{dc}}=\frac{U}{2}\sum_i [n_i^2-(n_i^0)^2]-\frac{1}{4}\sum_{i,\lambda} I_{\lambda}m_{i \lambda}^2
\end{equation}

 $n_i$ and $m_i$ are respectively  the charge and the spin moment of site $i$, $n_i^0$  the valence charge, $U$ is the Coulomb integral and $ I_{\lambda}$  the Stoner parameter of orbital $\lambda$  ($\lambda=s,p,d$ etc..). In transition metals $d$ orbitals are the one bearing the magnetism and the amplitude of magnetization is controled by the amplitude of  $ I_{d}$ (the exact value of $ I_{s}$ and $ I_{p}$ has a minor effect on the total magnetization but in practice we took $I_s=I_p=I_d/10$). 

The hopping and overlap integrals as well as onsite terms of $H_{\text{TB}}$ are fitted on ab initio datas (bandstructure and total energy). Local charge neutrality is controled by the amplitude of the Coulomb energy $U$ which in practice is taken equal to 20 $eV$. The value of the Stoner parameter $I_d$  is determined by reproducing ab-initio datas of the spin magnetization of bulk systems as a function of the lattice constants. The optimal $I_d$ value is the one that compares the best to ab-initio calculations. In the following we took $I_d$ = 0.88 $eV$. The spin-orbit constant $\xi_d$ is also determined by comparison with ab-initio bandstructure and we found that 60 $meV$ is a very good estimate for Iron.

\subsection{Force Theorem: FT}

The Force Theorem\cite{Weinert1985} has been used in various contexts. In studies of magnetic materials  it has mainly been used for the calculation of magnetocrystalline anisotropy\cite{Daalderop1990} or for the determination of exchange coupling in magnetic multilayers\cite{Mathon1997}. In this section we will illustrate its principle in a simple magnetic pure $d$ band orthogonal TB model of a monatomic system. The  total energy reads:

\begin{equation}
E_{\text{tot}}=\sum_{\alpha} f_{\alpha}  \langle \alpha |H| \alpha \rangle -\frac{U}{2}\sum_i (n_i^2-(n_i^0)^2) +\frac{I}{4}\sum_i m_i^2
\end{equation}

where the Hamiltonian $H$ is made of two terms:

 \begin{equation} 
H=H_{\text TB} +\sum_{i,\sigma,\lambda} |i,\lambda,\sigma \rangle \big [U  (n_i n_i^0)- \frac{I}{2} m_i \sigma \big] \langle i,\lambda,\sigma |
\end{equation}
The total energy obtained from this formula is caclulated by self-consistent loop on the charge and magnetic moment. Indeed the onsite terms of the Hamiltonian are renormalized by a quantity $\varepsilon_{i,\lambda,\sigma}=U(n_i-n_i^0) n_i-\frac{I}{2}m_i \sigma$  which depends itself on the local charge and magnetic moment. 

 Let us now consider the effect of a perturbative external potential $\delta V_{\text ext}$ which in our case will be the spin-orbit coupling. This external potential will induce a total potential variation $\delta V=\delta V_{\text ext} + \delta V_{\text ind}$ where $ \delta V_{\text ind}$ is the potential variation provoked by the modification of on site levels in the perturbed system. Within our model $ \delta V_{\text ind}$ is simply related to the variation $\delta n_i$ and $\delta m_i$ of the charge and magnetic moment thus,

\begin{equation}
 \delta V_{\text ind}=\sum_{i,\sigma,\lambda}   |i,\sigma,\lambda \rangle \big[U \delta n_i -\frac{I}{2}\delta m_i \big ] \langle i,\sigma,\lambda|
\end{equation}

The variation of the band energy due to  $ \delta V_{\text ind}$ can be straighforwardly calculated  from  first order pertubration expansion\cite{comment}:

\begin{equation}
\sum_{\alpha} f_{\alpha}  \langle \alpha | \delta V_{\text ind}| \alpha \rangle= U \sum_i n_i \delta n_i -\frac{I}{2}\sum_{i} m_i\delta m_i
\end{equation}

This variation is exactly compensated (to linear order) by the one of the double counting term and therefore the change of the total energy is equal to the change of band energy induced by the external potential only,  leading to the so-called force theorem:

\begin{equation}
\Delta E_{\text tot} \approx \Delta E_{\text{b}}^{\text{FT}}= \Delta \Big[\sum_{\alpha}  f_{\alpha}  \epsilon_{\alpha} \Big]
\end{equation}

Where $\Delta E_{\text{b}}^{\text{FT}}$ is the variation of the non-self-consistent band energy.   The great advantage of this formulation is obviously that self-consistency  effects can (and should)  be ignored. In this context, the total energy variation  induced by a change of the external potential from $ \delta V_{\text ext}^1$ to $ \delta V_{\text ext}^2$ (corresponding for instance to a change of the spin-orbit coupling matrices between two spin orientations $1$ and $2$) can be written:

\begin{equation}
\Delta E_{\text b}^{\text{FT}} \approx  \int^{E_F^1}En_1(E)dE-\int^{E_F^2}En_2(E)dE 
\end{equation}

$n_1(E)$ and $n_2(E)$ being the density of states and  $E_F^1$, $E_F^2$ the Fermi levels  of the configurations $1$ and $2$ respectively. The Fermi levels are determined by the condition on the  total number of electrons $N$  in the system:

\begin{equation}
N=\int^{E_F^1}n_1(E)dE=\int^{E_F^2}n_2(E)dE
\end{equation}

\subsection{ Grand canonical Force Theorem: FT$_{\text{gc}}$}

In the previous derivation of the Force Theorem it is important to note that the band energy is a summation of the eigenvalues over the occupied states (at fixed number of electrons) . Therefore a small variation of Fermi energy is expected with respect to the non perturbed system as follows:
 
\begin{equation}
E_F^1=E_F+\delta_1 \quad \text{and} \quad E_F^2=E_F+\delta_2
\end{equation}

 At linear order the variation of band energy can be written

\begin{equation}
\Delta E_{\text{b}}^{\text{FT}}=\int^{E_F}E\Delta n(E)dE+E_F(\delta_1 n_1(E_F)-\delta_2 n_2(E_F))
\end{equation}

using the conservation of the total number of electrons it comes that

\begin{equation}
\Delta E_{\text{b}}^{\text{FT}} \approx \Delta E_{\text{b}}^{\text{FT}_{\text{gc}}}=\int^{E_F}(E-E_F)\Delta n(E)dE 
\end{equation}

We will denote  FT$_{\text{gc}}$ this alternative formulation of the Force Theorem in the rest of the paper. 
The  FT$_{\text{gc}}$ formulation seems very similar to the standard FT formulation, but it leads to very different "space" partition of the energy. The underlying reason  is to be found in the type of statistical ensemble: canonical for FT and grand-canonical for FT$_{\text{gc}}$.  The grand-canonical  ensemble  for which the "good" variable is the Fermi energy (and not the total number of electrons) is better suited for a spatial partition of the energy\cite{Ducastelle_book}. For example the Gibbs construction\cite{MCDDS_book} to define properly surface quantities is based on a grand-canonical ensemble. Within this approach the suitable potential is the so-called grand-potential $\Omega=E_{\text b}-E_F N$. This formalism can be generalized at finite temperature\cite{Ducastelle_book,Cinal1997a,comment2}. Since the first-order variation of the Helmholtz Free Energy $F=E_{\text b}-TS_e$ at constant electron-number is equal to the fisrt order variation of the grand-potential at constant chemical potential the FT and FT$_{\text{gc}}$ formulation are equivalent in terms of variation of total energy. However the spatial repartition of energy could be very different within these two approaches.

\section{Magnetocrystalline anisotropy energy of Fe$(001)$ and Fe$(110)$  slabs }
\label{Sec:MAEslabs}

In this section we will present results on the MAE of  thin layers of Iron. In the first part (\ref{Sec.total_MAE}) we will discuss the validity of the various approximations presented in the methodological section. In particular we will justify the Force Theorem. We will also compare our results with ab-initio calculations proving the quality of our TB model. Sec. \ref{Sec.FTvsFTgc} will be devoted to the comparison of FT and FT$_{\text{gc}}$ formulations with respect to the layer resolved MAE and we will analyze  the surface anisotropy energy. Finally the Bruno formula will be discussed.

\subsection{MAE of Fe$(001)$ slabs: validity of the Force Theorem }
\label{Sec.total_MAE}

The MAE is defined as the change of total energy $E_{tot}$ associated to a change in the direction of the magnetization $\langle$ $\vec{S}$ $\rangle$ for a fixed position of atom. In the case of  a full self-consistent calculation including spin-orbit coupling, the MAE is defined as the energy difference
\begin{math}
\Delta E=E^{\perp}_{tot}-E^{\parallel}_{tot}
\end{math},
where $\perp$ and $\parallel$ refer to a magnetization where all atomic spins are pointing in a direction perpendicular or parallel to the surface respectively. The MAE is therefore the result of two independant self-consistent calculations which have to fulfill an extremely stringent condition for convergency since MAE are typically below meV. In systems containing "light" atoms like Iron for which the spin-orbit coupling constant $\xi$ is modest (60meV) it is expected that the Force Theorem should apply very well. Within this approximation the MAE is given by the difference of the band energy but ignoring any self-consistent effect. This type of calculation is performed in three steps: i) Collinear self-consistent calculation without SOC for which the density matrix is diagonal in spin space ii) Global rotation of the density matrix  to "prepare" it in the right spin direction iii) Non-collinear non-self-consistent calculation including SOC.  We have  performed  a series of calculations for ultrathin $(001)$  Iron layers  of various thicknesses ranging from one to twenty atomic layers, within the full-scf  and FT approaches. The lattice parameter of $a=2.85$~\AA~ was everywhere used and no atomic relaxation was considered. The convergency of the calculations have been carefully checked, we found that  2500 $k_\parallel$ and  4900 $k_\parallel$ points in the first  Brillouin zone for calculations without and with SOC respectively were sufficient to obtain a precision below 10$^{-5}$eV. The MAE obtained by these two methods differ by less than 10$^{-5}$eV proving the validity of FT approach which will be used systematically in the rest of the paper. It should be noted that FT approach leads to a considerable savings in the computational cost since no self-consistency is needed, therefore only one diagonalization of the full Hamiltonian including SOC is sufficient.

In order to check the  accuracy of our tight-binding model we have also performed {\sl ab-initio} calculations using the Quantum-ESPRESSO (QE) package \cite{Giannozzi2009} based on Density Functional Theory (DFT).  Since no FT approach is yet implemented in QE all the calculations are self-consistent and spin-orbit coupling is included via fully-relativistic ultrasoft pseudopotentials. 
The generalized gradient approximation (GGA) for exchange-correlation potential in the Perdew, Burke, and Ernzerhof parametrization was employed.
To describe thin films we have used the so-called super-cell geometry separating the adjacent slabs by about 8 \AA~ in the $z$ direction (orthogonal to 
the surface) in order to avoid their unphysical interaction. Since the MAE is usually a tiny quantity, ranging from $\mu$$eV$ to $meV$, it requires a very precise determination of total energy, and the total energy difference among various spin directions is very sensitive to the convergence of computational parameters. We found that $40 \times 40$ $k$-point mesh in the two-dimensional Brillouin zone was sufficient to obtain a well-converged MAE for $(001)$ Iron slabs . A Methfessel Paxton broadening scheme with 0.05 $eV$ broadening width was used with plane wave kinetic energy cut-offs of 30 Ry and 300 Ry for the wave functions and for the charge density, respectively.  Fig. 1 shows the total MAE as a function of the number layers of Fe $(001)$ slabs.  A good agreement is obtained between TB and {\sl ab initio} calculations which  proves once again the efficiency and quality of our TB model. 

\subsection{Layer-resolved MAE: FT versus FT$_{\text{gc}}$}
 \label{Sec.FTvsFTgc}

In section \ref{Sec.total_MAE}  we have only considered variations of total energies but it is also very instructive to investigate the local density of energy. Let us write the MAE as a sum of atomic-like contribution within FT and FT$_{\text{gc}}$ approaches:
\begin{equation} \label{FT_atomic}
\Delta E_{\text{b}}^{\text{FT}}=\sum_i \Bigg[\int_{E_{min}}^{E_{F}^{\perp}}En^{i}_{\perp}(E)dE-\int_{E_{min}}^{E_{F}^{\parallel}}En^{i}_{\parallel}(E)dE \Bigg]
\end{equation}
\begin{equation} \label{FTgc_atomic}
\Delta E_{\text{b}}^{\text{FT}_{\text{gc}}}=\sum_i \Bigg[\int_{E_{min}}^{E_{F}}(E-E_{F})\Delta n^{i}(E)dE\Bigg]
\end{equation}
where $n^{i}_{\perp}(E)$ and  $n^{i}_{\parallel}(E)$  are the density of states on atom $i$ for perpendicular or in-plane magnetization direction, respectively, and $\Delta n^{i}(E)=n^{i}_{\perp}(E)-n^{i}_{\parallel}(E)$. $E_{\text F}^{\perp}$  $E_{\text F}^{\parallel}$ are the corresponding Fermi energies and $E_{F}$ is the Fermi level of the collinear self-consistent calculation without SOC.  

The layer-resolved  MAE calculated by FT and FT$_{\text{gc}}$ methods for Fe $(001)$ slab  of 100 layers is shown in Fig.\ref{fig:100N}b. The most striking result is the very large oscillating behaviour which persists very deeply into the bullk  for the FT method. In addition, the local MAE obviously does not converge toward the expected bulk value which in this case should be exactly zero (since the three cubic axis are equivalent). In contrast, the layer resolved MAE obtained from the FT$_{\text{gc}}$ method corresponds to the behaviour expected from  a proper local quantity, namely a dominant variation in the vicinity of the surface that attenuates rapidly when penetrating in the bulk. This is indeed the case since only the surface atomic layer is strongly perturbed. In fact there are slight oscillations over the five first outer layers and an almost perfect convergence towards the bulk value for deeper layers. It is then clear that  FT$_{\text{gc}}$  is the appropriate method to define a layer resolved MAE. Note, however, that the total MAE are almost strictly  indentical for FT and FT$_{\text{gc}}$. Finally, it is very interesting to point out a striking analogy that exists with  the simple one-dimensional free-electron model  discusses in the next section \ref{Sec:QW1D}.

It is also useful to note the relation between Eq. \ref{FT_atomic} and Eq. \ref{FTgc_atomic} in order to understand the difference between the two methods:
\begin{equation}
\Delta E_{\text{b,i}}^{\text{FT}_{\text{gc}}}=\Delta E_{\text{b,i}}^{\text{FT}}-E_{F}(N^{i}_{\perp}-N^{i}_{\parallel})
\end{equation}
where  $N^{i}_{\perp}$ and  $N^{i}_{\parallel}$ are the Muliken charges on atom $i$ for perpendicular or in-plane magnetization, respectively. When summed over all the atoms of the system the additionnal term, $E_{F}(N^{\perp}-N_{\parallel})$, disappears since the total number of electrons is preserved and we recover the equivalence between FT and FT$_{\text{gc}}$ for total energy differences.  This formula is quite instructive since it shows that the difference between FT and FT$_{\text{gc}}$ is related to the slight charge redistribution between the two magnetic configurations.  At the first sight it seems that FT and FT$_{\text{gc}}$ should lead to very similar decomposition of the energy since the local charge neutrality term is supposed to avoid charge transfers and therefore $\Delta N^i=N^{i}_{\perp}-N^{i}_{\parallel} \approx 0$, but one should bear in mind that the force theorem applies only if self-consistency effects are ignored and therefore larger charge redistributions may appear. They produce irrelevant (to magnetic anisotropy) contributions  $E_F\Delta N^i$  to the local anisotropy energy which should be substracted as it is accomplished in the FT$_{\text{gc}}$ approach.
In Fig.\ref{fig:100N}a  we show $\Delta N^i$ which indeed looks very similar in shape to the FT layer resolved MAE and, when substracted,
leads thus to well behaved FT$_{\text{gc}}$ layer resolved MAE curve.  

These arguments show that the local variation of band energy should be the same after a self-consistent calculation provided that the local charge neutality is achieved.
To check this point  we have determined the layer-resolved  MAE for a  $(001)$  slab of 20 Fe layer  with full SCF calculation and FT$_{\text{gc}}$ method. Note that in the case of the full-scf approach one should consider the variation of the total energy wich includes band energy as well as double counting terms. In our TB scheme the double counting terms can easily be decomposed as a sum of atomic contributions and  will participate to the local MAE. In Fig. \ref{fig:SCF_FT} the layer-resolved MAE obtained from the two methods are presented and an excellent agreement between them 
is indeed found.

Finally let us point out an argument which was originally discussed by Daalderop {\sl et. al}\cite{Daalderop1990}: If a common Fermi energy  is used for the two direction of magnetization within the FT formulation then an additional term $E_F\Delta N$ is erroneously contributing to  the total MAE.

\subsection{Didactic example: one-dimensional quantum well}
\label{Sec:QW1D}

To illustrate the difference between FT and  FT$_{\text{gc}}$ let us consider one of the simplest models, a one-dimensional 
free-electron gas bounded within a length $L$ by infinite barriers (Fig.~\ref{fig.free-elec-1d}). 
The normalized wave functions and the corresponding discretized eigenvalues are (atomic units in which $\hbar^2=2m=1$ are used):

\begin{equation}
\psi_k(z)=\sqrt{\frac{2}{L}} \sin kz \quad \epsilon_k= k^2\quad \text{with} \quad k=p \frac{\pi}{L} 
\end{equation}

where $p$ takes only positive integer values. For the unbounded electron-gas with periodic Born-Von Karman (BVK) boundary conditions:

\begin{equation}
\psi_k^{\text{BVK}}(z)=\sqrt{\frac{1}{L}} e^{ikz} \quad \epsilon_k= k^2 \quad \text{with} \quad k=2n \frac{\pi}{L}
\end{equation}

In that case $n$ take any postive or negative integer values including 0. 
In the continuum limit the excess energy due to the creation of two surfaces is given by:

\begin{equation}
\Delta E=2 \times \frac{L}{\pi}\Big[\int_0^{k_F+\delta k_F}  \epsilon_k dk -\int_{0}^{k_F}  \epsilon_k dk\Big],
\label{DE_tot}
\end{equation}

where the factor $2$ is due to  the spin degeneracy and $k_F=\frac{\pi N}{2L}$ ($N$ is the total number of electrons in the box of the length $L$) 
is the Fermi wave vector of the unbounded homogeous gas.
Since an electron at $k=0$ is not allowed in the case of quantum well, it should be instead placed on the next free level,
which leads to $\delta k_F= \frac{\pi}{2L}$ and thus $\Delta E=k_F^2=E_F$.  Local decomposition of $\Delta E$
is naturally achieved by weighting each energy eigenvalue in (\ref{DE_tot}) by the squared modulus of the corresponding wave 
function which results in:    

\begin{equation}
\label{e}
\Delta E(z)= -\frac{2}{\pi} \int_{0}^{k_F} k^2 \cos(2kz) dk + \frac{2k_F^2}{L} \sin^2(k_Fz) 
\end{equation}

Equivalently, a grand-canonical formulation gives:

\begin{equation}
\label{e'}
\Delta E_{\text{gc}}(z)= -\frac{2}{\pi} \int_{0}^{k_F} (k^2-k_F^2) \cos(2kz) dk  \Big]
\end{equation}

Simple integration leads to exact expressions for $\Delta E(z)$ and $\Delta E_{\text{gc}}(z)$:

\begin{eqnarray}
\Delta E_{\text{gc}}(z) &=& \frac{1}{\pi}\Big( \frac{\sin(2k_F z)}{2 (k_Fz)^3}- \frac{\cos(2k_F z)}{ (k_Fz)^2}\Big) E_F k_F \\
\Delta E(z) &=& \Delta E_{\text{gc}}(z) -  \frac{\sin(2k_F z)}{\pi z}E_F  +\frac{2\sin^2(k_F z)}{L}E_F \\
\end{eqnarray}

These expressions, illustrated  in Fig.~\ref{fig.free-elec-1d}, are quite instructive. 
Within the FT$_{\text{gc}}$ formulation the density of surface energy behaves like $1/z^2$  for large $z$. 
The case of the FT formulation is more tricky: it contains, in addition, a term slowly decaying as $1/z$ and a term which does not decay (for a given $L$) but tends to zero as $L$ goes to infinity. In fact, these two last terms are simply proportional to the surface excess electronic density:  

\begin{eqnarray}
\Delta \rho(z) &=& -\frac{\sin(2k_F z)}{\pi z} +\frac{2\sin^2(k_F z)}{L}
\end{eqnarray}

so that $\Delta E(z)=\Delta E_{\text{gc}}(z) + E_F  \Delta \rho(z)$. Therefore, we conclude that long-range Friedel oscillations in $\Delta \rho(z)$
are at the origin of slow convergence with $z$ observed for the FT $\Delta E(z)$ which is perfectly in line with our previous 
analysis of layer-resolved magnetic anisotropies as illustrated by the striking similiraties between Fig. \ref{fig:100N} and Fig. \ref{fig.free-elec-1d}.
 
\subsection{MAE: surface and bulk contributions}

From the discussion above  it is natural to define the surface magnetic anisotropy energy as the sum of contributions from five outer layers (from both sides of the slab)  obtained using the FT$_{\text{gc}}$ formulation. The contributions from other layers sum up to what we call a bulk MAE. 
In Fig. \ref{fig:surface-volumeMAE} we plot the evolution of the surface, bulk and total MAE  for both Fe$(001)$ and Fe$(110)$ slabs with respect to the total number of layers $N$ (from 15 to 100).  
Note that the bulk MAE value per atom can be obtained by dividing the total bulk value by $N-10$  bulk-like layers. 
Also the true surface MAE should be obtained by dividing the surface contribution presented in Fig.  \ref{fig:surface-volumeMAE}  by two since the slabs contain two surfaces.
Our calculations show that (001) and (110) Fe surfaces have very different qualitative behaviour, the total MAE is negative for Fe$(001)$ indicating an out-of-plane easy axis while it is in-plane for Fe$(110)$ since its MAE is positive. More interestingly, in the case of  Fe$(110)$, additional calculations have shown that the magnitude of the in-plane anisotropy is almost as large as the one obtained between in-plane and out-of-plane orientations. It is also important to mention that the amplitude of the oscillations, though do not change the sign of the MAE, can however be as large as 0.2meV for Fe$(001)$ and 0.1meV for Fe$(110)$ at least up to $N \sim 40$. 
In addition, the total MAE is essentially dominated by the surface contribution. However, the oscillatory behaviour at large thicknesses, particularly pronounced for Fe$(001)$, clearly originates from the bulk. This kind of oscillatory behaviour of the MAE has been observed experimentally \cite{Przybylski2012,Manna2013} and was  interpreted in terms of quantum well states. The latter are formed in the ferromagnetic films from occupied and unoccupied electronic states close to the Fermi level that contribute significantly to the MAE.

\subsection{Bruno formula}

To gain better understanding of MAE beyond bare numbers, investigating related quantities is helpful. The orbital moment is a quantity essentially related to the SOC and to the MAE in magnetic systems.  It is well known that the easy axis always corresponds to the direction where the orbital moment is the largest. 
These arguments can be made more quantitative. Patrick Bruno\cite{Bruno1989} has derived an interesting relation using second order pertubation theory (since the first order term vanishes) with respect to the SOC parameter\cite{comment3}. Provided that the exchange splitting is large enough compared to the $d$-electron bandwidth,  the MAE can be made proportional to the variation of the  orbital moments. More precisely:
\begin{equation}
\label{Eq:Bruno}
E_{b,\perp}- E_{b,\parallel}=-\frac{\xi}{4}(\langle M_{\perp}^{\text{orb}}\rangle -(\langle M_{\parallel}^{\text{orb}} \rangle )
\end{equation}

This formula is based on a perturbative expansion (and an additionnal approximation concerning spin-flip transitions) for which the reference system and also the Fermi level are those of the unperturbed system without SOC. It can be shown that this approach is compatible with a grand canonical ensemble description (see Ref.\onlinecite{Cinal1997a} for a detailed discussion about statistical ensemble and second order corrections in the context of magnetic anisotropy). This relation can be generalized to systems with several atoms per unit cells\cite{Cinal1994} and also be used to extract a layer resolved MAE\cite{Gimbert2012}. In Fig. \ref{fig:Bruno} the layer resolved MAE calculated by Eq. \ref{Eq:Bruno} and by the Force Theorem are plotted, we found that only the surface layers have a significant contribution, while contribution from inner layers rapidly converges to the bulk (zero) value  within the two approaches. However, note that the Bruno's model results in quite different total MAE compared to the FT approximation in the vicinity of the surface. One can say that there is a rather good qualitative agreement between the two approaches, however the Bruno's formula can significantly (and quantitatively) differ from the FT$_{\text{gc}}$ results.

\section{Isolated Fe nanoclusters}
\label{Sec:MAEnanoclusters}

Once having properly defined the atomically resolved MAE  and analyzed in detail $(001)$ and $(110)$ Iron surfaces, it is interesting to study the case of clusters.
There exists a vast body of research on the theoretical investigation of combined structural and magnetic properties of unsupported transition metal clusters, relatively fewer are devoted to the determination of their magnetic anisotropy. Moreover most of them are dealing with small particles containing  few atoms\cite{Pastor1995,Nicolas2006,Desjonqueres2007}, the case of large clusters is generally treated with empirical Neel-like models of anisotropy\cite{Jamet2004}. In this section we will present TB calculation of  two large nanoclusters with  facets of orientations $(001)$ and $(110)$. We will more specifically consider the case of a truncated pyramid of nanometer size (see inset on Fig.\ref{fig:truncated-pyramid}). This geometry was chosen since such nanocrystals can be obtained by epitaxial growth on SrTiO$_3$$(001)$ \cite{Silly2005} substrate. 
In a second part we will consider the corresponding truncated bipyramid  made of two truncated pyramids joined at their bases.

\subsection{Truncated pyramid}
The particular cluster that we investigated is made of 620 atoms, with 12x12 atom lower base and 5x5 atom upper face and contains 8 atomic layers.
Its length-to-height ratio,  1.14 is close to the experimental value of $1.20 \pm0.12$. 
In Fig. \ref{fig:truncated-pyramid} we present the variation of the grand-canonical  band energy with respect to the Euler polar angle  $\theta$  between the magnetization direction and the $z$ axis 
choosen to be perpendicular to its "roof" and base  of $(001)$ orientation (see inset). 
The azimuthal angle $\phi$ is kept zero so that the magnetization remains in the $xz$ plane. The easy axis is evidently along the $z$ and the magneto-crystalline anisotropy is of the order of 110 meV. 
We also checked the azimuthal anisiotropy but found an extremely flat energy landscape in the $xy$ plane with an amplitude of 3 meV, the hard axis being along the diagonal of the base. To get more insight into the origin of the anisotropy we have decomposed the band energy per atomic sites and analyzed different contributions: total surface, $(001)$ facets, perimeter of the base, etc. 
Summing local MAE over atomic sites in the outer shell of the nanocluster (dashed line), we almost recover the total magnetocrystalline anisotropy proving that only the outer shell (so called surface atoms) is participating to the overall anisotropy. A more detailed analysis  showed  only two significative contributions: 
i) low coordinated perimeter atoms of the base (red line) and 
ii) two $(001)$ facets, excluding perimeter atoms (blue line). 

Intrestingly, the perimeter atoms have the strongest anisotropy while, on the contrary,
the contribution from $(110)$ side facets is almost negligible (and, moreover, cancel each other because of their opposite orientations).

By counting the number of "implied" atoms (109 $(001)$ atoms and 44 perimeter atoms) it is possible to extract an average anisotropy per $(001)$ surface atom and per perimeter atom. One finds $0.56$ meV/atom and $0.90$ meV/atom for $(001)$ and perimeter atoms, respectively. This coresponds quite well to the expected anisotropy found for the Fe$(001)$ slabs.

\subsection{Truncated bipyramid}

We then consider another type of cluster: a truncated bipyramid (lower inset in Fig. \ref{fig:truncated-bipyramid}) made of 1096 atoms and obtained by attaching symmetrically to the 
previous truncated pyramid another one (with removed base plane) from below.
In Fig. \ref{fig:truncated-bipyramid} we have compared the total MAE of the two nanoclusters. 
Although the truncated bipyramid contains more atoms its anisotropy (15meV) is much  lower than in previous case. 
The explanation is quite straightforward from the previous analysis: the surface of the $(001)$ facets has been strongly reduced and, 
moreover, the perimeter atoms of the base have now more neigbours and no longer contribute so strongly to the total anisotropy. 
The latter comes from two small $(001)$ facets only. 
This argument works rather well: indeed, the number of atoms in $(001)$ facets is now 18 which gives an anisotropy of $18\times 0.56=10$meV, 
the value slightly smaller then the overall MAE, 15meV, with the missing contribution coming from perimeter atoms which were not taken 
into account.

\section{Conclusion}
\label{Sec:conclusion}
A comprehensive TB study of magnetocrystalline anisotropy energy of Fe$(001)$ and Fe$(110)$ slabs and nanoclusters has been presented. Due to small spin-orbit coupling constant, the Force Theorem  is valid  for Fe-based systems studied in this work. We have shown that a proper way to define the layer-resolved MAE should use the grand-canonical FT formulation instead of the standard FT, while the two approaches are equivalent for the total MAE. The prefered orientations for Fe$(001)$ and Fe$(110)$ slabs are out of plane and in-plane, respectively. For both slabs, the total MAE is dominated by surface contribution as expected. However, surface contribution converges more rapidly than bulk one with respect to the number of atomic layers of the slabs. The study of nanoclusters showed that the dominating MAE originates form the $(001)$  facets and especially from low coordinated perimeter atoms of the base of the pyramid. On the contrary, the contribution from $(110)$ side facets is almost negligible. In view of the results presented here, a study of the magnetic properties of nanoclusters deposited on a substrate (SrTiO$_3$ (001), Au or Cu etc ...) is rather promising since depending on the bonding between the substrate and the $(001)$ facets one could imagine to tune the magnetic anisotropy of  these nanoclusters. Finally Skomski {\sl et al}\cite{Skomski2007} showed that the shape of surface anisotropy could have consequences on the magnetization reversal of nanoparticles. Therefore, it is very likely that a detailed investgation of the spin dynamics of  nanometer size iron clusters could reveal such surface effects in the anisotropy.

\noindent{\bf Acknowledgement}

 The research leading to these results has received funding from the European Research Council under the European Union's Seventh Framework Programme (FP7/2007-2013) / ERC grant agreement n$^{\circ}$ 259297. This work was performed using HPC resources from GENCI-CINES (Grant Nos.  x2013096813).


\bibliographystyle{apsrev}
\bibliography{biblioMAE}

\begin{thebibliography}{30}
\bibitem[{com({\natexlab{a}})}]{comment}
\bibinfo{note}{The first order variation of an eigenvalue induced by a
  perturbative potential $\delta V$ is given by the formula $\delta
  \varepsilon_{\alpha}= \langle \alpha | \delta V| \alpha \rangle$. The band
  energy of the perturbed system is therefore $\displaystyle \sum_{\alpha} f(
  \varepsilon_{\alpha}+\delta \varepsilon_{\alpha})
  (\varepsilon_{\alpha}+\delta \varepsilon_{\alpha} )$. Using the conservation
  of the total number of electrons it comes that the band energy variation is
  equal to $\displaystyle \sum_{\alpha} f( \varepsilon_{\alpha}) \delta
  \varepsilon_{\alpha} $.}
\bibitem[{com({\natexlab{b}})}]{comment2}
\bibinfo{note}{The grand potential at finite temperature $T$ and chemical
  potential $\mu$ can be written in two alternative ways (thanks to an
  integration by parts) : $\displaystyle \Omega(T)=\int L(E)n(E)dE=-\int
  f(E)N(E)dE$. With $\displaystyle L(E)=-k_BT\log \{1+\exp[- (E-\mu) /k_BT] \}$
  and $\displaystyle N(E)=\int^E n(E')dE'$. Note that the derivative of $L(E)$
  is the Fermi function $f(E)=[1+\exp(E-\mu)/k_BT]$}.
\bibitem[{com({\natexlab{c}})}]{comment3}
\bibinfo{note}{The second order perturbation expansion at finite temperature is
  given by the well-known formula $\displaystyle \delta \Omega=\sum_{\alpha,
  \beta } |\langle \alpha |H_{\rm{SO}}|\beta\rangle|^2
  \frac{f_{\alpha}(1-f_{\beta})}{\varepsilon_{\alpha}-\varepsilon_{\beta}}$}.
\end{thebibliography}

\clearpage

\begin{figure}[!hbp]
\centering
\includegraphics[width=12cm]{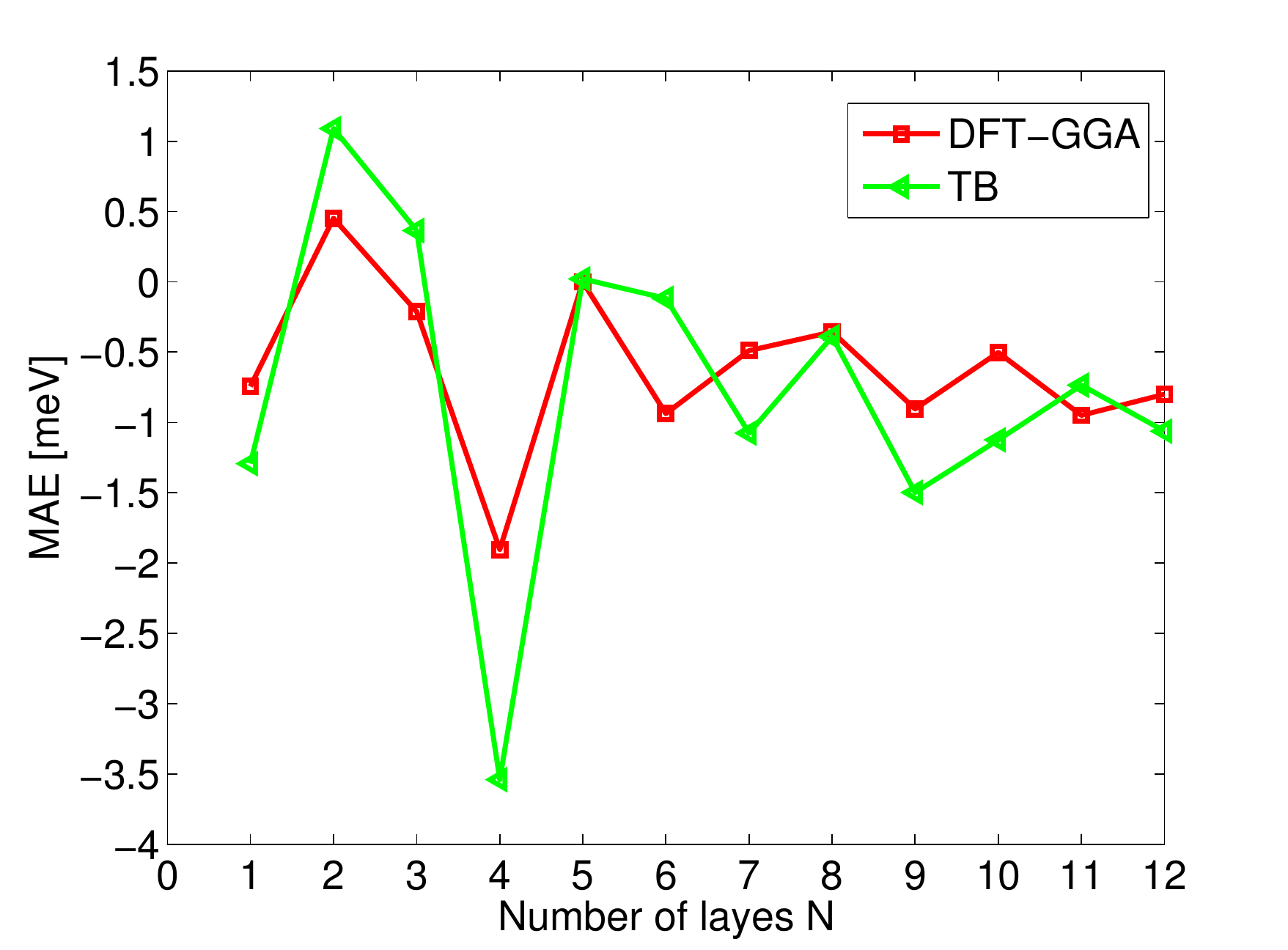}
\caption{\label{Ab-initio_TB}
Total MAE versus Fe film thickness  $N$ for  Febcc $(001)$ slabs.
TB calculation (in green) are compared with {\sl ab-initio} DFT-GGA calculations (in red).}
\end{figure}

\begin{figure}[!hbp]
\centering
\includegraphics[width=12cm]{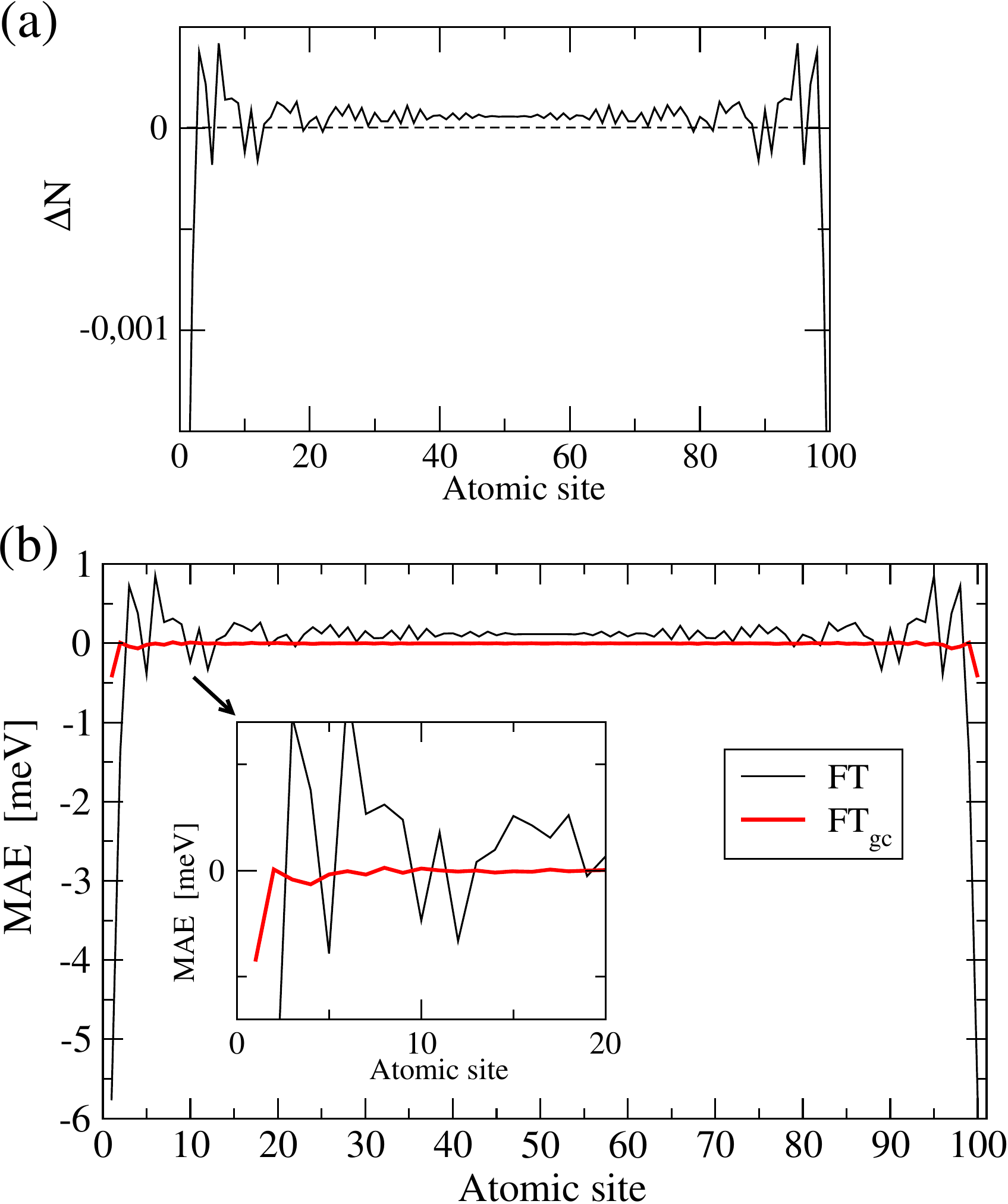}
\caption{\label{fig:100N}
a)  Variation of the charge difference $\Delta N_i=(N^{i}_{\perp}-N^{i}_{\parallel})$  between out of plane and in plane magnetic configurations (obtained after one diagonalization)  on successive atomic layers of a Fe$(001)$ slab  containing $N=100$ layers.
b) Layer resolved MAE of the Fe$(001)$ slab calculated with two different methods:  canonical FT (black lines) and grand canonical FT$_{\text{gc}}$ (red lines). 
The zoom over the first 20 layers is shown in the inset. }
\end{figure}

\begin{figure}[!hbp]
\centering
\includegraphics[width=12cm]{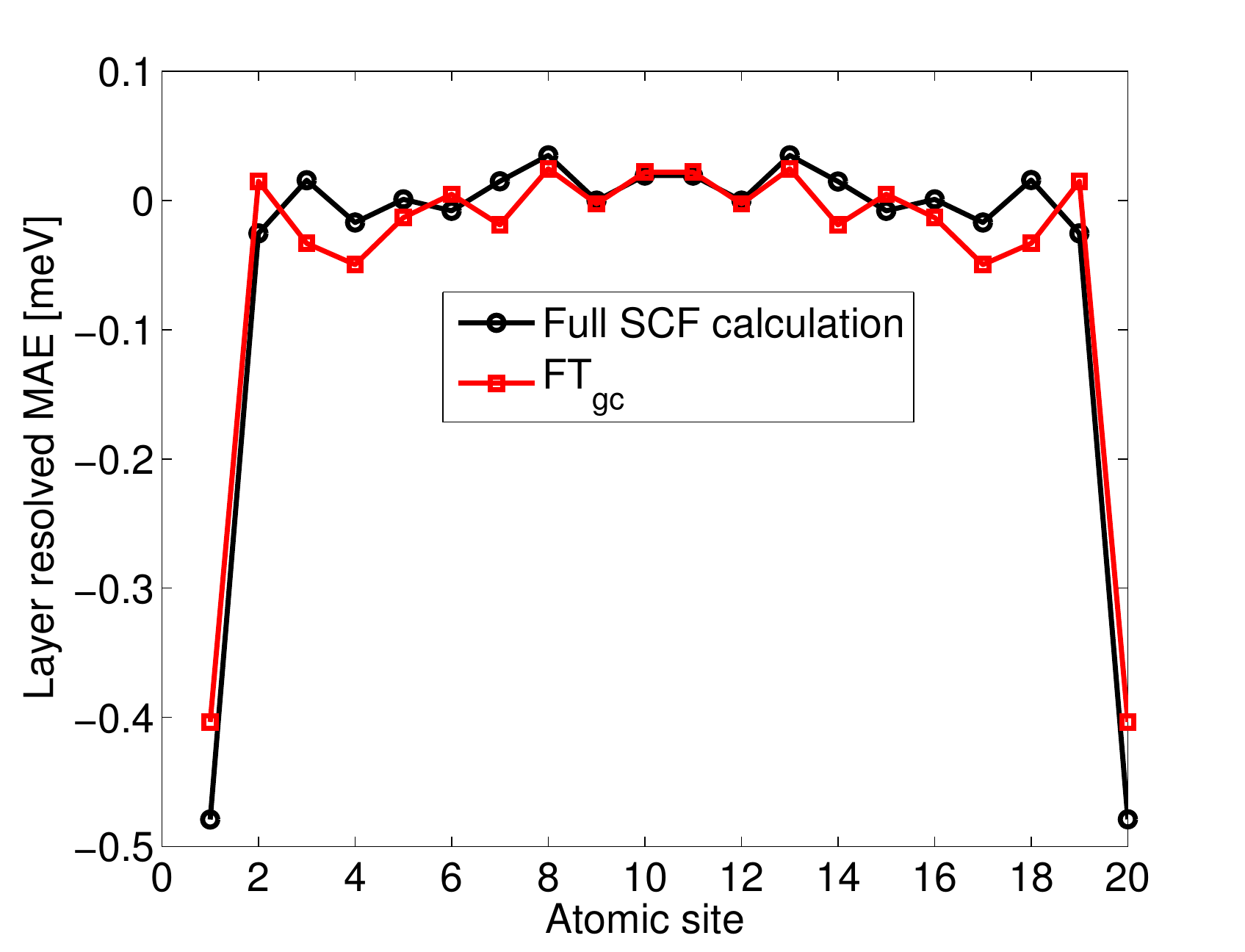}
\caption{\label{fig:SCF_FT}
Layer-resolved MAE of the Fe$ (001)$ slab with $N = 20$ layers calculated using the TB fully self-consistent calculation and FT$_{\text{gc}}$ approximation.
Very good agreement confirms a proper local decomposition of MAE provided by the grand canonical formulation.
}
\end{figure}

\begin{figure}[!hbp]
\centering
\includegraphics[width=12cm]{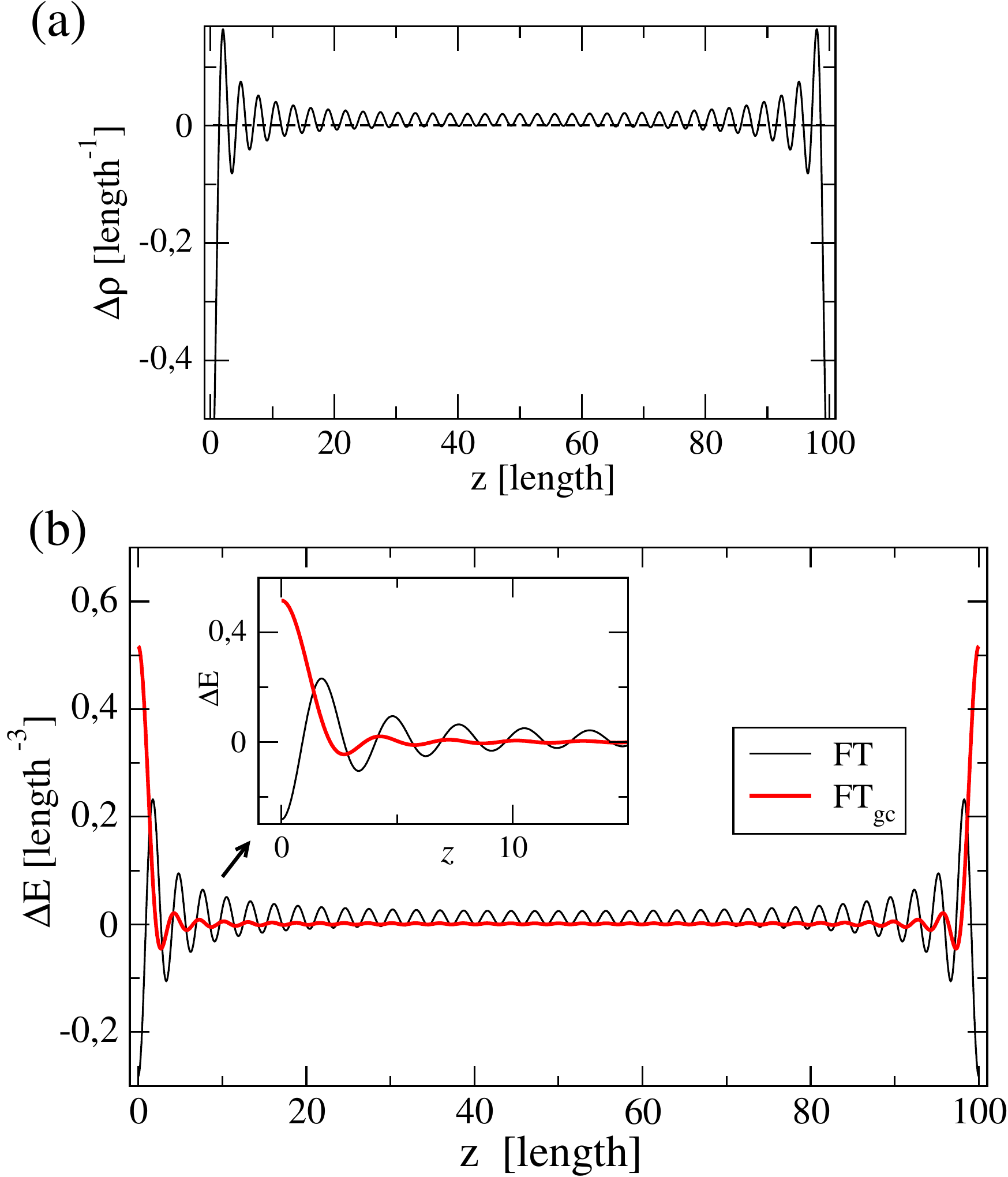}
\caption{\label{fig.free-elec-1d} Graphical representation of the functions $\Delta E(z)$, $\Delta E_{\text{gc}}(z)$,  
and $\Delta \rho (z)$ for a one-dimensional electron gas confined by infinite barriers in the box of the length $L$. 
The discretized calculations were done with the parameters $N=70$ (total number of electrons) and $L=100$.
}

\end{figure}

\begin{figure}[!hbp]
\centering
\includegraphics[width=12cm]{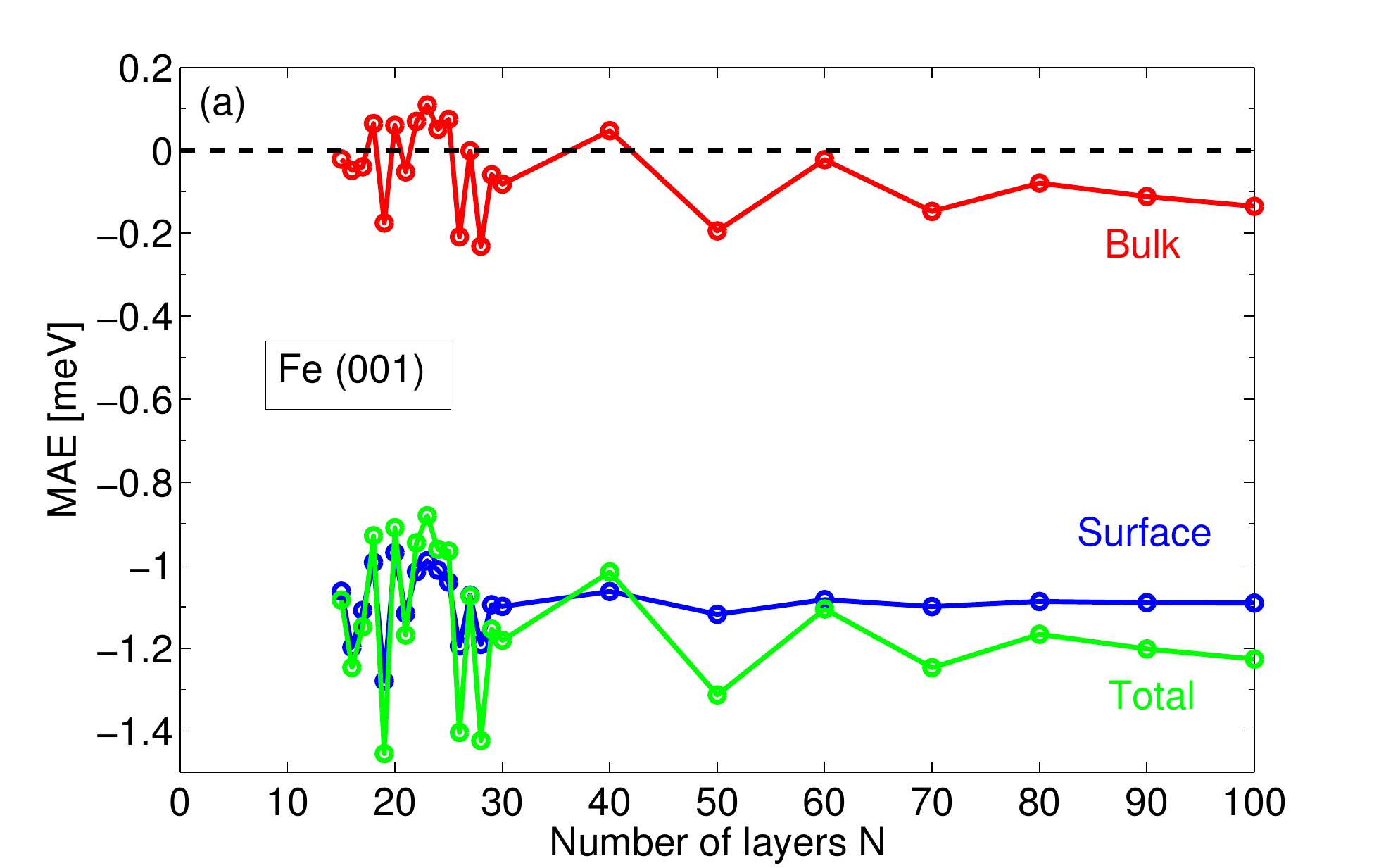}
\includegraphics[width=12cm]{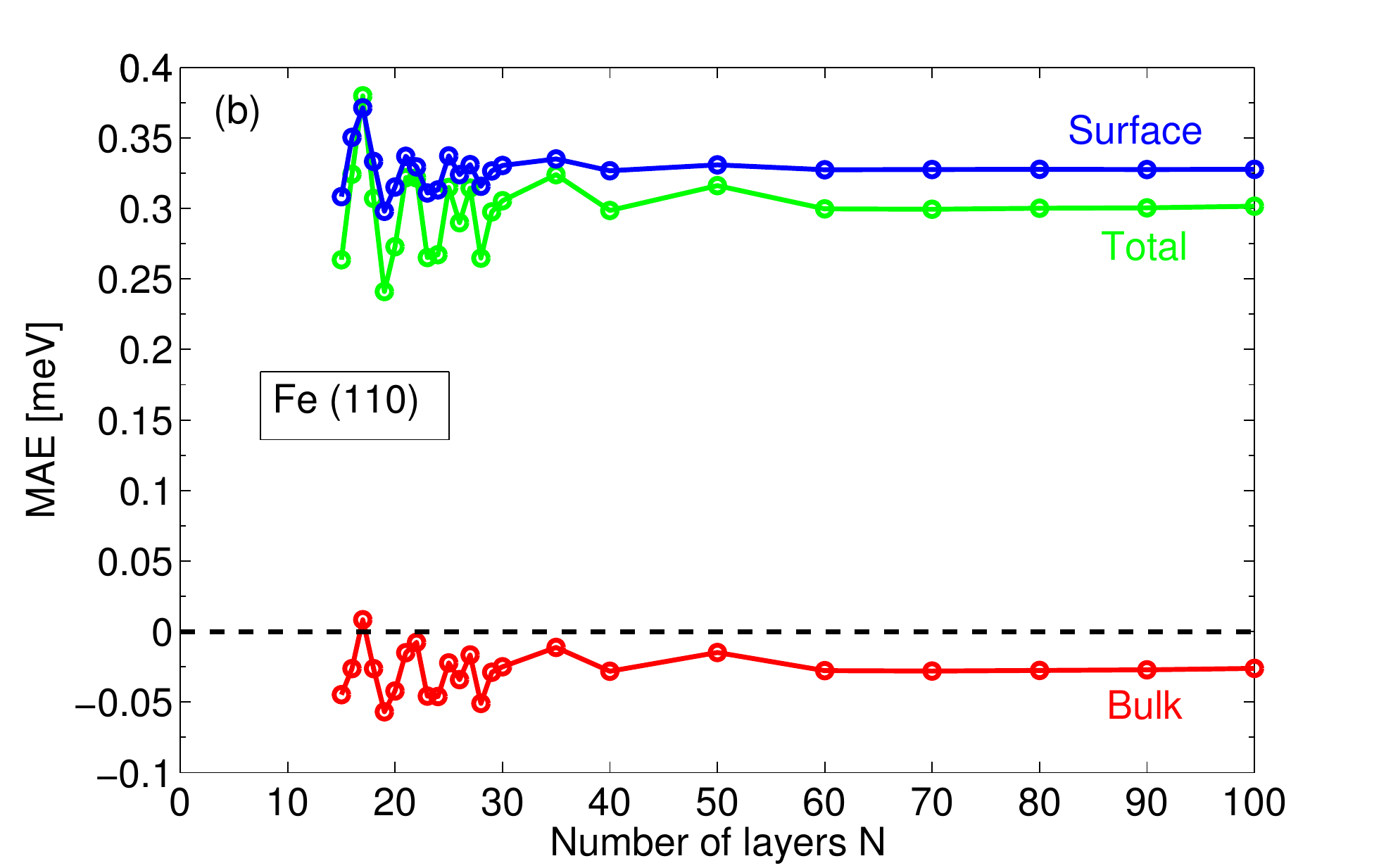}
\caption{\label{fig:surface-volumeMAE}
(Color online) Surface , bulk (see definitions in the text), and total MAE for Fe$(001)$ and Fe$(110)$  slabs as a function of the film thickness. The surface contribution is obtained by summing the layer resolved MAE over the 5 outer layers on each side of the slab. Consequently the true surface MAE can be obtained by dividing by 2 this quantity. Positive (negative) MAE values mean in (out of) plane easy axis direction.
The two different slab orientations have magnetic anisotropies of opposite sign.
}
\end{figure}

\begin{figure}[!hbp]
\centering
\includegraphics[width=12cm]{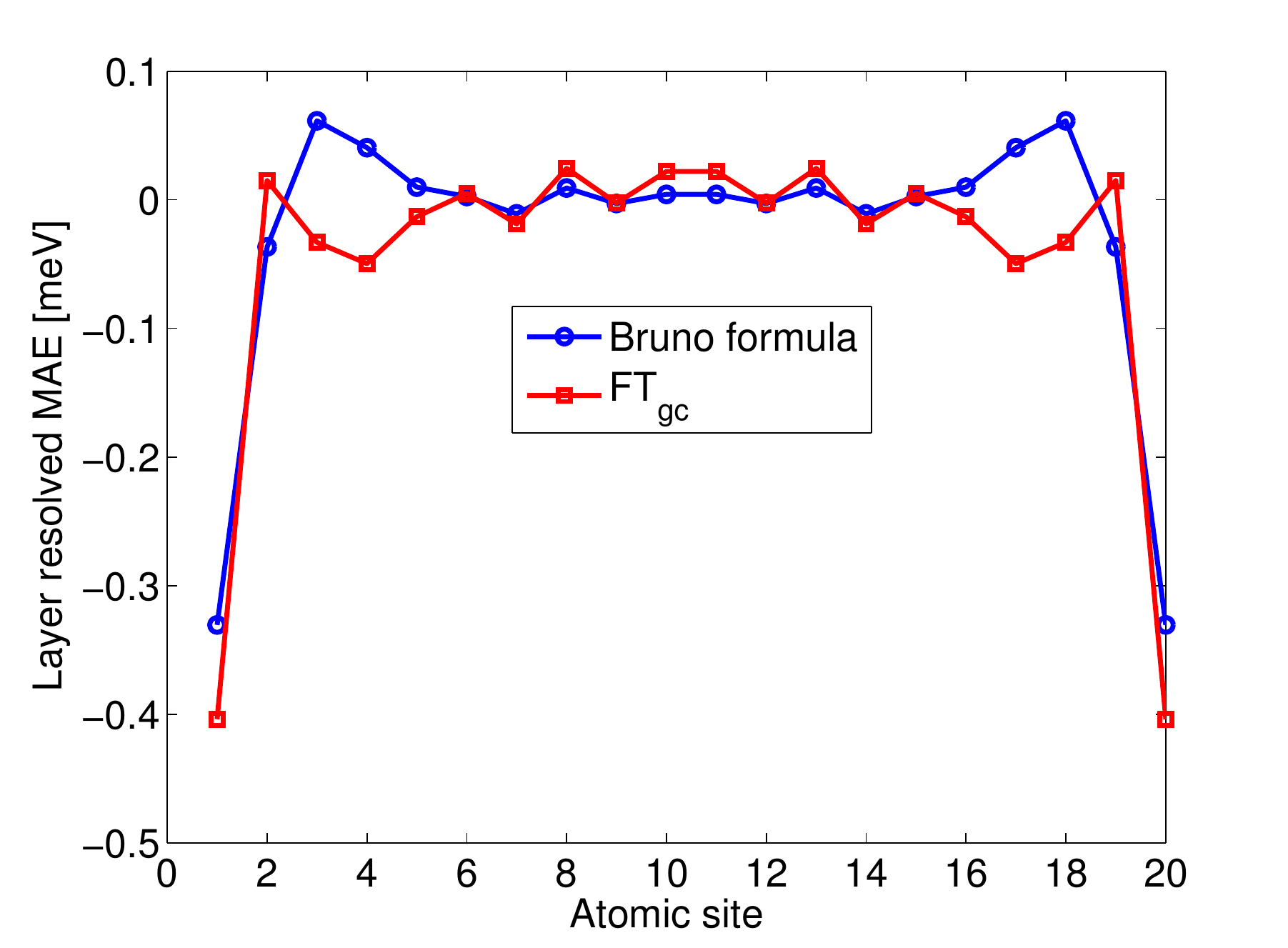}
\caption{\label{fig:Bruno}
Layer-resolved MAE of the Fe $(001)$ slab with $N = 20$ layers obtained from  FT$_{\text{gc}}$ (red line)  and the Bruno formula (blue line).  
The layer-resolved spin-projected orbital moments are obtained by self-consistent calculations.}
\end{figure}

\begin{figure}[!hbp]
\centering
\includegraphics[width=12cm]{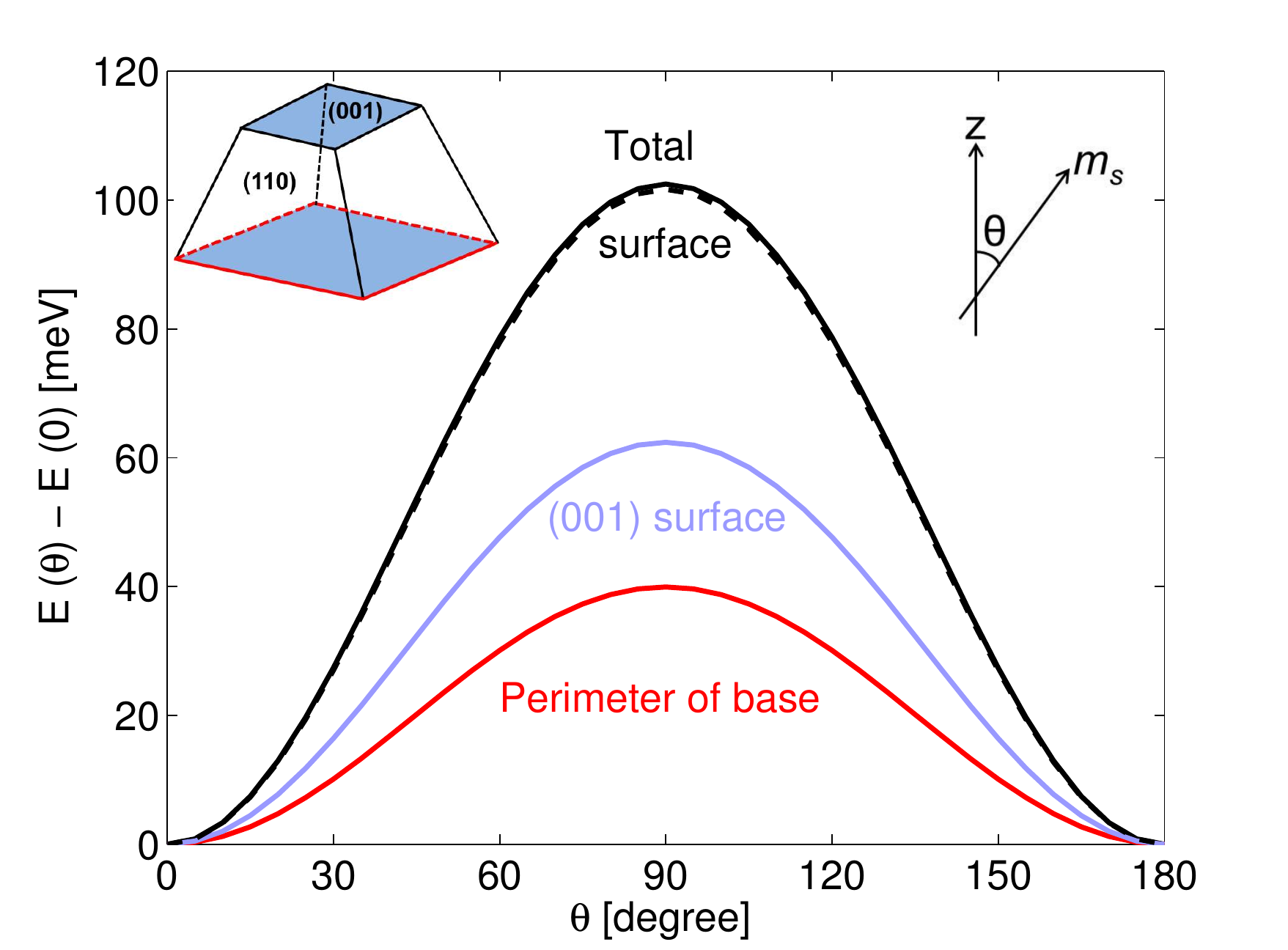}
\caption{\label{fig:truncated-pyramid}
(Color online) Magnetocrystalline anisotropy of a truncted pyramid with $N = 620$ atoms, as a function of the angle $\theta$ between the $z$ axis ($[001]$ direction) and the direction of the spin. Contributions from atoms of the two  $(001)$ facets (excluding perimeter atoms) and from perimeter atoms of the base are shown in blue and red lines, respectively.  The total MAE and the contribution from atoms of the outer shell  (surface)  are represented in full  and dashed black lines which are almost superposed. $E(\theta = 0)$ is taken as the zero of energy. Note that in all calculations the azimuthal angle $\phi$ is equal to zero.}
\end{figure}

\begin{figure}[!hbp]
\centering
\includegraphics[width=12cm]{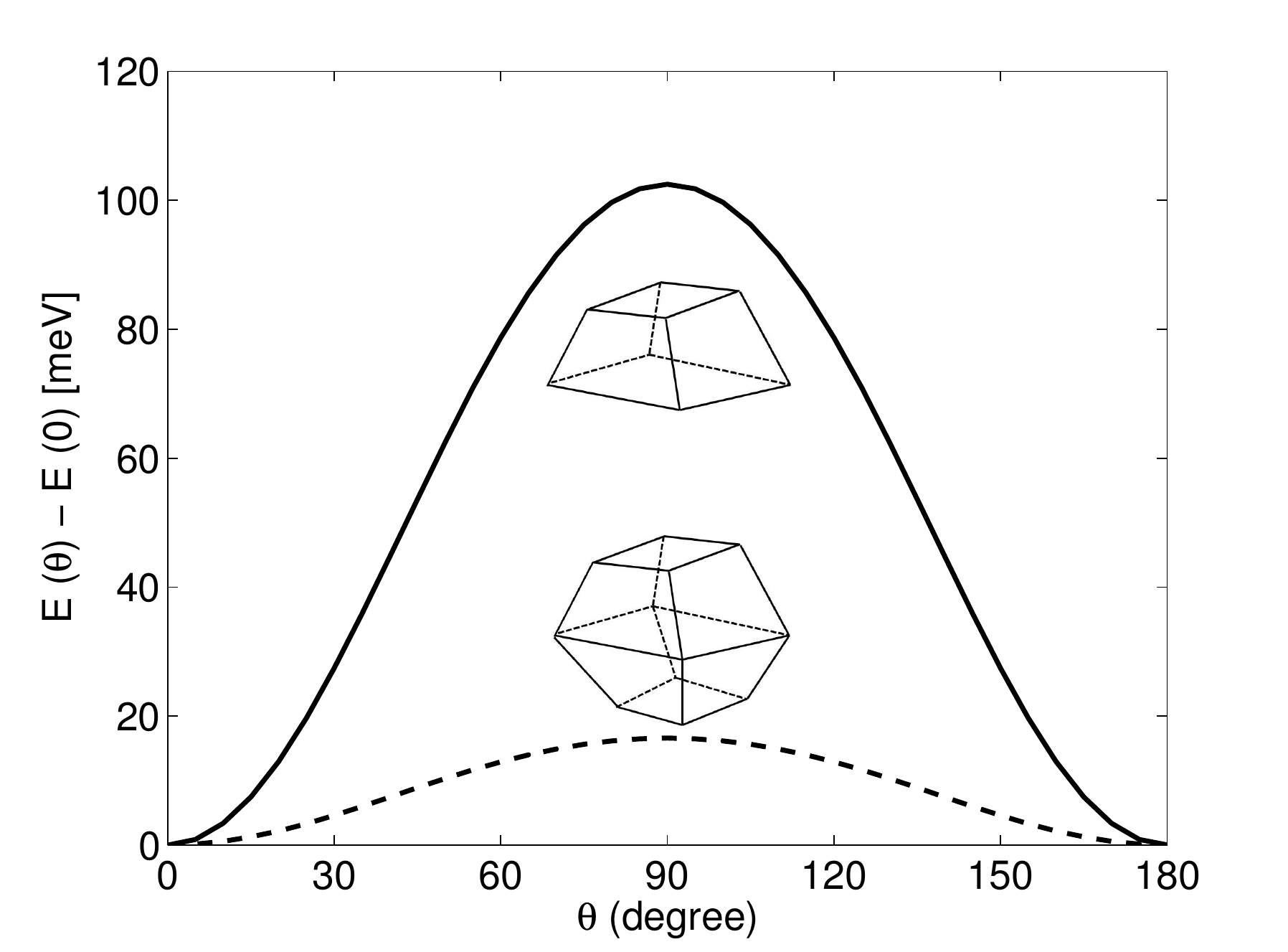}
\caption{\label{fig:truncated-bipyramid}
Total MAE as a function of angle $\theta$ for a truncated pyramid ($N = 620$) and a truncated bipyramid ($N = 1096$). For the latter,
the MAE is strongly reduced because of much smaller area of $(001)$ facets and strongly reduced anisotropy from perimeter atoms. }
\end{figure}

\end{document}